\documentclass[12pt]{article}
\usepackage{amsmath,amsfonts}
\usepackage[nosort]{cite}
\usepackage{graphicx}

\makeatletter\@addtoreset{equation}{section}\makeatother

\def\bH {\mathbb{H}}
\def\bR {\mathbb{R}}

\newcommand{\vev}[1]{{\left< {#1} \right>}}

\newcommand{\Tr}{{\rm Tr\,}}

\newcommand{\cL}{{\cal L}}
\newcommand{\cN}{{\cal N}}
\newcommand{\cP}{{\cal P}}

\renewcommand{\title}[1]{\vbox{\center\LARGE{#1}}\vspace{3mm}}
\renewcommand{\author}[1]{\vbox{\center#1}\vspace{3mm}}
\newcommand{\address}[1]{\vbox{\center\em#1}}
\newcommand{\email}[1]{\vbox{\center\tt#1}\vspace{3mm}}

\begin{document}

\begin{titlepage}
\begin{center}
\hfill {\tt HU-EP-06/46}\\
\hfill {\tt YITP-SB-06-56}\\
\hfill{\tt Imperial/TP/06/RR/01}\\
\hfill {\tt hep-th/0612168}\\
\vskip .3cm

\title{On the D3-brane description \\ of some $1/4$ BPS Wilson loops
}

\author{Nadav Drukker$^{1,a}$,
Simone Giombi$^{2,b}$,
Riccardo Ricci$^{3,4,c}$,
Diego Trancanelli$^{2,d}$}

\address{$^1$Humboldt-Universit\"at zu Berlin, Institut f\"ur Physik,\\
Newtonstra{\ss}e 15, D-12489 Berlin, Germany\\
\medskip
$^2$C. N. Yang Institute for Theoretical Physics,\\
State University of New York at Stony Brook\\
Stony Brook, NY 11794-3840, USA\\
\medskip
$^3$ Theoretical Physics Group, Blackett Laboratory,\\
Imperial College, London, SW7 2AZ, U.K.\\
\medskip
$^4$ The Institute for Mathematical Sciences, \\
Imperial College, London, SW7 2PG, U.K.}

\email{$^a$drukker@physik.hu-berlin.de,
$^b$sgiombi@max2.physics.sunysb.edu,
$^c$r.ricci@imperial.ac.uk,
$^d$dtrancan@max2.physics.sunysb.edu}

\end{center}

\abstract{
\noindent
Recently it has been proposed that Wilson loops in high-dimensional 
representations in $\cN=4$ supersymmetric Yang-Mills theory (or multiply 
wrapped loops) are described by D-branes in $AdS_5\times S^5$, rather
than by fundamental strings. Thus far explicit D3-brane solutions have 
been only found in the
case of the half-BPS circle or line. Here we present D3-brane solutions
describing some $1/4$ BPS loops. In one case, where the loop is
conjectured to be given by a Gaussian matrix model, the action of the
brane correctly reproduces the expectation value of the Wilson loop
including all $1/N$ corrections at large $\lambda$. As in the 
corresponding string solution, here too we find two classical solutions,
one stable and one not. The unstable one contributes exponentially small
corrections that agree with the matrix model calculation.
}

\vfill

\end{titlepage}

\tableofcontents

\section{Introduction}

In the $AdS$/CFT correspondence \cite{Maldacena:1997re} Wilson
loops are usually described by macroscopic fundamental strings that end
along the loop at the holographic boundary of the space
\cite{Rey,Maldacena-wl}. This realizes the string picture of gauge
theories, albeit in a conformal, not confining setting. In string theory
some objects can metamorphose into others and it was indeed realized in
\cite{Drukker:2005kx} that a multiply wrapped Wilson loop
(or a loop in a high-dimensional representation) is better described by
a D-brane rather than by a large number of coincident strings\footnote{
Already in \cite{Rey} D3-branes were proposed as a possible 
holographic description of Wilson loops.}.

Initially a D3-brane configuration was found and later also a D5-brane
configuration. In both cases they corresponded to a $1/2$ BPS Wilson
loop, either a straight line or a circle. The D3-brane describes
the Wilson loop in the symmetric representation (which, in the strong 
coupling regime, seems to be
dominated by the single-trace multiply wrapped loop)
whereas the D5-brane gives the loop in the antisymmetric representation
\cite{Hartnoll:2006hr,Yamaguchi:2006tq,Gomis:2006sb,
Okuyama:2006jc,Hartnoll:2006is,Gomis:2006im}.

Since the initial description of the half-BPS loop in terms of D3-branes,
such solutions were not found in any other system. It turned out to be
simpler to find D5-brane solutions, and some non-supersymmetric
examples were studied
\cite{Hartnoll:2006hr,Hartnoll:2006ib,Armoni:2006ux}.

In this paper we find some systems where it is possible to find
 solutions for the D3-branes starting from first-order equations derived from
the supersymmetry conditions. All the examples we present are Wilson loop
operators which preserve $1/4$ of the supersymmetry generators.
First we consider the system of a
straight $1/2$ BPS Wilson loop with the insertion of two $1/2$ BPS
local operators such that the combined system preserves $1/4$ of the
supercharges. If there was only the Wilson loop, the D3-brane would have
been the one of \cite{Drukker:2005kx}, while if only the local operators
were present, that would have involved the original giant gravitons
\cite{McGreevy:2000cw,Grisaru:2000zn,Hashimoto:2000zp}.

This combined system of a Wilson loop and a local operator was presented in
\cite{Drukker:2006xg}, where it was shown to be supersymmetric and the relevant
string solution was found. In Section~2 we present the D3-brane solution
preserving the same supersymmetries, and interpolating between the
``giant Wilson loop'' of \cite{Drukker:2005kx} near the boundary and
a giant graviton in the center of $AdS_5$.

The second system, which will be described in Section~3 involves
circular Wilson loops which couple to three of the $\cN=4$ scalars. This system,
first presented in \cite{Drukker:2005cu} (as a generalization of an
example in \cite{Zarembo:2002an}) and studied further in
\cite{Drukker:2006ga}, also preserves $1/4$ of the supersymmetries, 
but with a different combination of generators than in the previous
example. At two loop order in the perturbative expansion the interacting graphs
(in the Feynman gauge) cancel, which led to the conjecture that only
ladder/rainbow diagrams contribute to these operators. All those
diagrams combine nicely into a matrix model which was then compared
with the string calculation in $AdS_5\times S^5$. As we shall review
in Section~3 below, the results agreed including a subleading term,
a world-sheet instanton, which matched a correction to the asymptotic
expansion of the matrix model at strong coupling.

The calculation using D3-branes is applicable for a Wilson loop in 
a symmetric representation whose rank $k$ is of order $N$. At large 
$\lambda$ the analog observable in the matrix model agrees with the 
single-trace multiply wrapped loop which is given by a function of 
the ratio $k/N$ and thus the D3-brane calculation captures non-planar
corrections to the usual string calculation. We are able to compare the
matrix model and the D-brane calculation for arbitrary $k/N$ and find
an agreement and a check of the aforementioned conjecture to all orders
in $1/N$.

The two systems presented in Sections~2 and~3 are quite different.
We combined them in the same paper since they both preserve eight
supercharges, and some of the technical details of the calculations are
similar. Also, the system in Section~2 serves as a good warm-up
exercise to the more interesting system in Section~3. Because of the
differences between the two systems, more details on the two setups
will be given in each of those sections.

\section{Wilson loop with insertions}

\subsection{Setup}
We consider here a Wilson loop operator in $\cN=4$ supersymmetric
Yang-Mills theory on $S^3\times\bR$, where the line is the time
direction with Lorentzian signature. The loop will be comprised of one
line in the time direction along a point on $S^3$, and another line
going in the opposite direction at the antipodal point. In addition we
will include the insertion of local operators at the infinite past and infinite
future.

Under the exponential map (after Wick-rotation), the space is mapped to
flat $\bR^4$ and the two lines to a single line through the origin with
the local insertions at the origin and at infinity. Without the insertions
this Wilson loop preserves half the supersymmetries of the vacuum, and
we will consider the local insertions to also be half-BPS, $Z^J$, where
$Z=\Phi_1+i\Phi_2$ is a complex scalar field (if $Z^J$ is at the origin
the charge has to be absorbed by $\bar Z^J$ at infinity). Note that the
insertions are not gauge invariant, since they are not traced over, and transform
in the adjoint representation of the gauge group. The entire configuration
is nonetheless gauge invariant, because of the presence of the Wilson loop. This
guarantees that in the string picture the charge generated by the local
operator is carried by the open string (or D-brane) representing the
Wilson loop, and not by another supergravity field.

Recall that the half-BPS Wilson loop contains also a coupling to one
of the scalars. For the combined system to be $1/4$ BPS, this scalar has
to be orthogonal to $Z$ and $\bar Z$, so we take it to be $\Phi_3$.
Formally we can write the Wilson loop as
\begin{equation}
W_{Z^J}=\Tr\cP \left[Z^J(-\infty)
e^{i\int_{-\infty}^\infty (A_t(t,0)+\Phi_3(t,0))dt}
\bar Z^J(\infty)
e^{i\int_\infty^{-\infty} (A_t(t,\pi)+\Phi_3(t,\pi))dt}\right]\,.
\label{WL-insertion}
\end{equation}
The arguments of $A_t$ and $\Phi_3$ are the time and the two points
on $S^3$ given by an angle at $0$ and $\pi$.

\subsection{String solution}

This Wilson loop was studied in \cite{Drukker:2006xg} as were some
non-supersymmetric generalizations of it and they were related to a certain
spin-chain system. There it was proven that (\ref{WL-insertion}), which
was the pseudo-vacuum of the spin-chain system, is supersymmetric. Also
the string solution describing it at large $J$ and large 't Hooft coupling
$\lambda$ was given. We review it here.

Take the following metric for $AdS_5\times S^2$ (the other directions
on $S^5$ do not play any role and we do not write them explicitly)
\begin{equation}
\frac{ds^2}{L^2}=
-\cosh^2\rho\,dt^2+d\rho^2+\sinh^2\rho\left(d\chi^2
+\sin^2\chi(d\vartheta^2+\sin^2\vartheta\,d\varphi^2)\right) + 
d\theta^2+\sin^2\theta d\phi^2\,. \label{metric-string1}
\end{equation}
$L$ is the radius of curvature related to the 't Hooft coupling and the string
tension by $L^4=\lambda\,\alpha'^2$. The Wilson loop should reach the
boundary at $\chi=0$ and $\chi=\pi$. At those points it should approach
$\theta=0$ on the $S^2$, which is the direction corresponding to
$\Phi_3$. In the bulk the string should rotate around this sphere carrying
the angular momentum related to $Z^J$.

The solution to the string equations of motion which satisfies these
conditions is
\begin{equation}
\phi=t\,,\qquad
\sin\theta=\frac{1}{\cosh\rho}\,. \label{stringsol1}
\end{equation}
There are two parts to the string: at $\chi=0$ and at $\chi=\pi$. They
are continuously connected to each other beyond $\rho=0$.
For a full derivation of the solution see \cite{Drukker:2006xg}.

Some interesting issues arise when studying the analog system in
Euclidean signature. Those were discussed in \cite{Miwa:2006vd}.

\subsubsection{Supersymmetry analysis}

A precise counting of the supersymmetries preserved by the string solution (\ref{stringsol1}) was performed in \cite{Drukker:2006xg}. Here we briefly review that computation. 

The number of supersymmetries 
preserved by the string is equal to the number of independent solutions to the 
equation $\Gamma\epsilon=\epsilon$. The $\kappa$-symmetry projector $\Gamma$ is given by
\begin{equation}
\Gamma=\frac{1}{\sqrt{-\det g}}\,
\partial_t x^\mu\partial_\rho x^\nu
\gamma_\mu\gamma_\nu K\,, \label{stringgamma1}
\end{equation}
where $g$ is the induced metric on the world-sheet parameterized by $t$ and 
$\rho$, $K$ acts by complex conjugation, and 
$\gamma_\mu=e_\mu^a\Gamma_a$ with $\Gamma_a$ constant tangent space 
gamma-matrices. The dependence of the Killing spinors $\epsilon$ 
on the relevant coordinates of the metric (\ref{metric-string1}) is
\begin{equation}
\epsilon=e^{-\frac{i}{2}\rho\,\Gamma_\star\Gamma_1}
e^{-\frac{i}{2}t\,\Gamma_\star\Gamma_0}
e^{-\frac{i}{2}\theta\,\Gamma_\star\Gamma_5}
e^{\frac{1}{2}\phi\,\Gamma_{56}}\epsilon_0\,,
\label{Killing-string-sec2}
\end{equation}
where $\Gamma_\star=\Gamma^0\Gamma^1\Gamma^2\Gamma^3\Gamma^4$ is
 the product of all gamma-matrices in the $AdS_5$ directions and 
 $\epsilon_0$ is any constant chiral complex 16-component spinor.
The spinors $\epsilon$ solve the Killing equation
\begin{equation}
\left(\partial_{\mu} + \frac{1}{4} \omega_{\mu}^{ab}\Gamma_{ab}+\frac{i}{2L}\Gamma_{\star}\gamma_{\mu} \right)\epsilon=0\, .
\end{equation}

Inserting the solution (\ref{stringsol1}) into the expression (\ref{stringgamma1}) it is easy to see that
$\Gamma$ does not depend on $t$. The only place where $t$ appears 
is in the exponent of 
the Killing spinors. Since the projection equation has to hold for all $t$ and $\rho$ we eliminate this dependence by imposing the condition
\begin{equation}
\Gamma_\star\Gamma_{056}\epsilon_0=i\epsilon_0\,,
\label{SUSY-cond-1}
\end{equation}
so that the Killing spinors become
\begin{equation}
\epsilon=e^{-\frac{i}{2}\rho\,\Gamma_\star\Gamma_1}
e^{-\frac{i}{2}\theta\,\Gamma_\star\Gamma_5}
\epsilon_0\,.
\label{Killing-2}
\end{equation}
After some manipulation the action of the projector can be written as
\begin{equation}
\Gamma\epsilon=
-e^{-\frac{i}{2}\rho\,\Gamma_\star\Gamma_1}e^{
-\frac{i}{2}\theta\,\Gamma_\star\Gamma_5}
\,\Gamma_{01}K\epsilon_0\,,
\end{equation}
so the projector equation is solved by all constant spinors satisfying
\begin{equation}
\Gamma_{01}K\epsilon_0= - \epsilon_0\,.
\label{SUSY-cond-2}
\end{equation}
It is easy to verify that the two conditions (\ref{SUSY-cond-1}) and (\ref{SUSY-cond-2}) are consistent with each-other,  so there are eight linearly independent real solutions to this equation. Thus 
the string solution preserves $1/4$ of the supersymmetries.

\subsection{D3-brane solution}

We look now for the D3-brane solution associated to this Wilson loops with insertions.
The loop is in the time direction, as reviewed above, and preserves an
$SO(3)\times SO(3)$ symmetry, the
first being part of the $AdS_5$ isometry and the other coming from the $S^5$.
It is convenient to use the metric (\ref{metric-string1}) and fix a static gauge where 
$t$, $\rho$, $\vartheta$ and $\varphi$ are the world-volume
coordinates on the D3-brane. The ansatz is then
\begin{equation}
\chi=\chi(\rho)\,,\qquad
\theta=\theta(\rho)\,,\qquad
\phi=t\,. \label{ansatz1}
\end{equation}

The brane action consists of a Dirac-Born-Infeld (DBI) part and of a Wess-Zumino (WZ) term, which captures the coupling to the background RR form
\begin{equation}
S
=T_{D3} \int e^{-\Phi} \sqrt{-\det (g + 2\pi\alpha' F)} - T_{D3} \int P[C_4]\, , 
\end{equation}
where $T_{D3}=\frac{N}{2\pi^2 L^4}$ is the brane tension and $P[C_4]$ 
denotes the pullback of the 4-form to the brane world-volume. The solution 
should include a non-zero electric field, carrying $k$ units of flux 
associated to the Wilson loop. Because of the symmetry of the system,
it will be in the direction $F_{t \rho}(\rho)$. 

With the ansatz above the DBI action reads (in the following we absorb a 
factor of $2\pi\alpha'/L^2$ in the definition of $F_{t\rho}$)
\begin{equation}
\begin{aligned}
S_{DBI} &=
\frac{2 N}{\pi}  \int dt \, d\rho \, \sinh^2\rho \sin^2\chi 
\sqrt{(\cosh^2\rho-\sin^2\theta)
(1+\sinh^2\rho\,\chi'^2+\theta'^2)- 
F_{t\rho}^2}\,,
\end{aligned}
\label{DBI}
\end{equation}
whereas the WZ term is given by
\begin{equation}
S_{WZ}=\frac{2 N}{\pi} \int dt \, d\rho \, \sinh^4\rho \sin^2\chi \,  \chi'  \,,  
\label{WZ}
\end{equation}
and the relative sign between these two terms in the action is positive. 
In these formulas the $'$ denotes a derivative with respect to $\rho$.

It is rather complicated to solve the equations of motion coming from this action. Instead of trying to do this, we write down the supersymmetry equations derived from requiring $\kappa$-symmetry. These are first-order rather than second-order and can be integrated easily. 

The $\kappa$-symmetry projector associated with the D3-brane embedding is (see for example \cite{Skenderis:2002vf})
\begin{equation}
\Gamma=\cL_{DBI}^{-1}\left(\Gamma_{(4)}+
L^2F_{t\rho}\Gamma_{(2)}K\right)I\,,
\end{equation}
where $K$ acts by complex conjugation, $I$ by multiplication
by $-i$, and 
\begin{equation}
\begin{aligned}
\Gamma_{(4)}&=\partial_t x^{\mu} \partial_{\rho} x^{\nu}\partial_{\vartheta} x^{\xi}\partial_{\varphi} x^{\zeta} \gamma_{\mu} \gamma_{\nu}\gamma_{\xi}\gamma_{\zeta}= 
(\gamma_t+\gamma_\phi)
(\gamma_\rho+\chi'\,\gamma_\chi+\theta'\,\gamma_\theta)\gamma_\vartheta\gamma_\varphi\,,\\
\Gamma_{(2)}&=\partial_{\vartheta} x^{\mu} \partial_{\varphi} x^{\nu} \gamma_{\mu} \gamma_{\nu}=
\gamma_\vartheta \gamma_\varphi\,, 
\end{aligned}
\end{equation}
with $\gamma_\mu=e_\mu^a\Gamma_a$.
Using the vielbeins 
\begin{equation}
\begin{gathered}
e^0=L\cosh\rho\,dt\,,\qquad
e^1=L\,d\rho\,,\qquad
e^2=L\sinh\rho\,d\chi\,,\qquad
\\
e^3=L\sinh\rho\sin\chi\,d\vartheta\,,\qquad
e^4=L\sinh\rho\sin\chi\sin\vartheta\,d\varphi\,,
\\
e^5=L\,d\theta\,,\qquad
e^6=L\sin\theta\,d\phi\,,
\end{gathered}
\end{equation}
and the ansatz (\ref{ansatz1}), the projectors $\Gamma_{(4)}$ and $\Gamma_{(2)}$ can be explicitly written as
\begin{equation}
\begin{aligned}
\Gamma_{(4)}&=L^2(\cosh\rho\Gamma_0+\sin\theta\Gamma_6)
(\Gamma_1+\sinh\rho\,\chi'\,\Gamma_2+\theta'\,\Gamma_5)\,
\Gamma_{(2)}\,, \\
\Gamma_{(2)}&=L^2\sinh^2\rho\sin^2\chi\sin\vartheta\,\Gamma_{34}\,.
\end{aligned}
\end{equation}
Adding the dependence on the other coordinates into 
(\ref{Killing-string-sec2}), 
the Killing spinors for the metric (\ref{metric-string1}) are
\begin{equation}
\epsilon=e^{-\frac{i}{2}\rho\,\Gamma_\star\Gamma_1}
e^{-\frac{i}{2}t\,\Gamma_\star\Gamma_0}
e^{\frac{1}{2}\chi\,\Gamma_{12}}
e^{\frac{1}{2}\vartheta\,\Gamma_{23}}
e^{\frac{1}{2}\varphi\,\Gamma_{34}}
e^{-\frac{i}{2}\theta\,\Gamma_\star\Gamma_5}
e^{\frac{1}{2}\phi\,\Gamma_{56}}\epsilon_0\,. \label{killing1}
\end{equation}

From the supersymmetry analysis in the string case,  we know that the constant spinors
$\epsilon_0$ satisfy the conditions (\ref{SUSY-cond-1}) and (\ref{SUSY-cond-2})
\begin{equation}
K\epsilon_0=-\Gamma_{01}\epsilon_0\,,\qquad
\Gamma_6\epsilon_0=-i\Gamma_{12345}\epsilon_0\,.\label{constraint1}
\end{equation}
Plugging $\phi=t$ in the expression (\ref{killing1}) and using the second constraint in the equation above, the Killing spinors may be rewritten as
\begin{equation}
\epsilon=e^{-\frac{i}{2}\rho\,\Gamma_\star\Gamma_1}e^{ -\frac{i}{2}\theta\,\Gamma_\star\Gamma_5 }
e^{\frac{1}{2}\chi\,\Gamma_{12}}
M \epsilon_0\, ,
\end{equation}
where
\begin{equation}
M=e^{\frac{1}{2}\vartheta\,\Gamma_{23}} e^{\frac{1}{2}\varphi\,\Gamma_{34}}.
\end{equation}
The differential equations we are looking for will come from considering 
the projector equation
\begin{equation}
\Gamma \epsilon = \epsilon\,.
\end{equation}
To simplify it, we move the matrix 
$e^{-\frac{i}{2}\rho\,\Gamma_\star\Gamma_1}
e^{ -\frac{i}{2}\theta\,\Gamma_\star\Gamma_5} 
e^{\frac{1}{2}\chi\,\Gamma_{12}}$ 
to the left of the projector $\Gamma$,  using some gamma-matrix algebra and 
applying the constraints (\ref{constraint1}) (note that $\epsilon_0$ and 
$M \epsilon_0$ satisfy the same constraints). In this way we get a set of 
8 differential equations in $\theta$, $\chi$ and $F_{t\rho}$ 
(on the left we indicate the gamma-matrix structure the equations come from)
\begin{equation}
\begin{array}{lcl}
\Gamma_{0345}:
&&0=F_{t\rho}\sinh\rho\cos\chi\sin\theta
-\theta'(\cosh^2\rho-\sin^2\theta)\,\\
\Gamma_{\star} \Gamma_5:
&&0=F_{t\rho}\sinh\rho\sin\chi\sin\theta
-\chi'\sinh^2\rho\sin\theta\cos\theta\,\\
\Gamma_{0234}:
&&0=(\cosh^2\rho-\sin^2\theta)\sin\chi+ \chi'\cosh\rho\sinh\rho\cos\chi\cos^2\theta\,\\
\Gamma_{12}:
&&0=F_{t\rho}\sinh\rho\cos\chi\cos\theta+\theta'\sin\theta\cos\theta+\cosh\rho\sinh\rho\,\\
\Gamma_{15}:
&&0=\chi'\cosh\rho\sinh\rho\cos\chi\sin\theta\cos\theta-\theta'\cosh\rho\sinh\rho\sin\chi
+\sin\chi\sin\theta\cos\theta\,\\
\Gamma_{25}:
&&0=F_{t\rho}\cosh\rho\sin\theta+\cos\chi\sin\theta\cos\theta-
\chi'\cosh\rho\sinh\rho\sin\chi\sin\theta\cos\theta\\
&& \hskip .7cm -\theta'\cosh\rho\sinh\rho\cos\chi\,\\
\Gamma_{0134}:
&&0=F_{t\rho}\cosh\rho\cos\theta+(\cosh^2\rho-\sin^2\theta)\cos\chi-
\chi'\cosh\rho\sinh\rho\sin\chi\cos^2\theta\,\\
1:
&&1=-L^4 \cL_{DBI}^{-1}\sinh^2\rho\sin^2\chi\sin\vartheta\left(F_{t\rho}\sinh\rho\sin\chi\cos\theta+
\chi'\sinh^2\rho\sin^2\theta\right)\,
\end{array}
\end{equation}
One can solve for $ \theta',\,  \chi' $ and $ F_{t\rho}$ using for instance
the first three equations. Once these are solved, the
remaining five are automatically satisfied. The first three equations give
\begin{equation}
\begin{aligned}
\theta'&= - \tan\theta \tanh\rho\,,\\
\chi'\,\cot\chi&=-\frac{\cosh^2\rho-\sin^2\theta}
{\cosh\rho\sinh\rho\cos^2\theta}\,,\\
F_{t\rho}&=-\frac{\cosh^2\rho-\sin^2\theta}
{\cosh\rho\cos\theta\cos\chi}\,.
\end{aligned}
\label{three-eqns}
\end{equation}
The solution to the first equation is
\begin{equation}
\sin\theta = \frac{C_1}{\cosh\rho}. \label{sol1theta}
\end{equation}
The integration constant $C_1$ is related (in a complicated way) 
to the amount of angular momentum carried by the brane. 
After plugging the solution for $\theta$ into the expressions for
$\chi'$ and $F_{t\rho}$,
we find
\begin{equation}
\chi'\,\cot\chi=
- \frac{\cosh^4\rho-C_1^2}{\sinh\rho \cosh\rho(\cosh^2\rho-C_1^2)}\, ,
\end{equation}
which is solved by
\begin{equation}
\sin\chi = C_2 \, \frac{\coth\rho}{\sqrt{\cosh^2\rho-C_1^2}}\, , \label{sol1chi}
\end{equation}
with $C_2$  a second integration constant. Finally the electric field is 
\begin{equation}
F_{t\rho}=-\frac{\cosh^4\rho-C_1^2}
{\cosh^2\rho\sqrt{\cosh^2\rho-C_1^2-C_2^2\coth^2\rho}}\, . \label{sol1F}
\end{equation}

Plugging the BPS equations (\ref{three-eqns}) into the DBI action (\ref{DBI}), the
square root simplifies to 
\begin{equation}
\sqrt{(\cosh^2\rho-\sin^2\theta)
(1+\sinh^2\rho\,\chi'^2+\theta'^2)-F_{t\rho}^2}
=(\cosh^2\rho-\sin^2\theta)
\frac{\tanh\rho \tan\chi}{\cos^2\theta}\,.
\end{equation}
It is then straightforward to check that the solutions (\ref{sol1theta}), (\ref{sol1chi}) and (\ref{sol1F}) satisfy the brane equations of motion stemming from (\ref{DBI}) and (\ref{WZ}).

\subsubsection{Conserved charges}
The solution has two integration constants $C_1$ and $C_2$ which are related to
the two conserved charges carried by the brane: the rank of the symmetric 
representation (or the number of windings of the Wilson loop) $k$, and 
the angular momentum $J$ around the $S^2$ in the  $S^5$.

The first charge $k$ is the conjugate momentum to the gauge field after integrating over $\vartheta$ and $ \varphi$ 
\begin{equation}
k=\Pi= \frac{2\pi\alpha'}{L^2} \, T_{D3} \int d\vartheta\,d\varphi\,
\frac{\delta\cL}{\delta  F_{t\rho}}
=\frac{4N}{\sqrt\lambda}C_2\,.
\end{equation}
Then
\begin{equation}
C_2=\frac{k\sqrt\lambda}{4N}\equiv \kappa\,.
\end{equation}
If we take $\kappa\rightarrow 0$ we recover the string solution (\ref{stringsol1}). Notice that the electric field $F_{t\rho}$ does not vanish in this limit.

The other conserved charge carried by the brane is the angular momentum $J$
\begin{equation}
\begin{aligned}
J&=2\, T_{D3} \int d\vartheta\, d\varphi\, d\rho\,
\frac{\delta \cL}{\delta\dot\phi}\\
&=-\frac{4N}{\pi}\int d\rho\,
\frac{\sinh^2\rho\sin^2\chi \sin^2\theta\, (1+\sinh^2\rho \, \chi'^2 +\theta'^2)}{\sqrt{(\cosh^2\rho-\sin^2\theta)
(1+\sinh^2\rho\,\chi'^2+\theta'^2)- 
F_{t\rho}^2}
} \, . 
\label{J}
\end{aligned}
\end{equation}
Here the range of the $\rho$ integral is $[\text{arccosh}\,C_1,\,\infty)$, 
which like in the 
case of the string, covers only half the world-volume, with $\chi<\pi/2$. 
A multiplicative factor of 2 was included to account for the other branch 
with $\chi>\pi/2$.

Plugging the explicit solutions in this expression it is easy to see that $J\rightarrow 0$ when $C_1\rightarrow 0$. In this limit the brane does not rotate along the $S^2$ and the solution reduces to
\begin{equation}
\begin{aligned}
\sin\theta&=0\,,\\
\sin\chi\,\sinh\rho&=\kappa\,,\\
F_{t\rho}&=-\frac{\cosh\rho}{\cos\chi}\,,
\end{aligned}
\end{equation}
which is, after a conformal transformation, the same as the 1/2 BPS brane of \cite{Drukker:2005kx}.

The energy gets contributions from the DBI action and the Wess-Zumino term
\begin{equation}
\begin{aligned}
E_{DBI}&=2\, T_{D3} \int d\vartheta\, d\varphi\, d\rho\,
\frac{\delta \cL_{DBI}}{\delta\dot t}\\
&=\frac{4N}{\pi}\int d\rho\,
\frac{\sinh^2\rho\sin^2\chi \cosh^2\rho\,  (1+\sinh^2\rho \, \chi'^2 +\theta'^2)}{\sqrt{(\cosh^2\rho-\sin^2\theta)
(1+\sinh^2\rho\,\chi'^2+\theta'^2)-
F_{t\rho}^2}\,,
} \\
E_{WZ}&=2 \, T_{D3}\int d\vartheta\, d\varphi\, d\rho\,
\frac{\delta \cL_{WZ}}{\delta\dot t}
=\frac{4N}{\pi}\int d\rho\,\sinh^4\rho\sin^2\chi\, \chi' \,.
\label{E-bulk}
\end{aligned}
\end{equation}
In addition one has to add a total derivative term, which serves as a Legendre 
transform from the gauge field coordinate to the conjugate momentum 
$\Pi$, which is the correct canonical variable in this 
problem \cite{Drukker:2005kx}. This is
\begin{equation}
E_{L.T.}=\frac{2L^2}{2\pi\alpha'} \int d\rho \, \Pi\,F_{t\rho}
=-\frac{4N}{\pi}\kappa \int d\rho\, \frac{\cosh^2\rho-\sin^2\theta}{\cosh\rho \cos\theta \cos\chi}\, .
\label{E-boundary}
\end{equation}

Plugging  the BPS equations in the formulas (\ref{J}), (\ref{E-bulk}) 
and (\ref{E-boundary}) above, one can see that
\begin{equation}
E_{DBI}+E_{WZ}+E_{L.T.}+J= \frac{4N}{\pi}\int d\rho\,
\frac{\cosh^2\rho-\sin^2\theta}{\cosh\rho \cos\chi} \left[ \sinh\rho\sin\chi-\frac{\kappa}{\cos\theta} \right].
\end{equation}
Using the explicit solution it is easy to check that the term in square brackets vanishes, so
we get $E=-J=|J|$.

\section{Wilson loop wrapping a circle on $S^5$}

\subsection{Setup}
In this section we shall look at a family of circular Wilson loops that
couple to three of the six scalars of the $\mathcal{N}=4$ multiplet. These operators 
were presented in
\cite{Drukker:2005cu} and studied in detail in \cite{Drukker:2006ga}.
As for the loop in the previous section, the $S^5$ part
of the gravity calculation will reduce to an $S^2$ subspace. The difference will be that  here
the couplings to the scalars are smeared around the loop and not localized
at two points.

While the Wilson loop follows a curve on the boundary of $AdS_5$ parameterized by
\begin{equation}
x^1=R\cos s\,,\qquad
x^2=R\sin s\,,
\end{equation}
the scalar to which it couples is given by the linear combination
\begin{equation}
\Phi(s)=\Phi_3\cos\theta_0+\sin\theta_0(
\Phi_1\cos s+\Phi_2\sin s)\,,
\end{equation}
with an arbitrary fixed parameter $\theta_0$.
The loop may be written (in Euclidean signature) as
\begin{equation}
W_{\theta_0}
=\Tr\cP \exp \left[\oint \left (i A_\mu(s) \dot x^\mu+
|\dot x|\Phi(s)\right)ds\right]\,.
\label{WL-circle}
\end{equation}
In the special case of $\theta_0=0$ this is the usual half-BPS circle, while
for $\theta_0=\pi/2$ this is a special case of the supersymmetric Wilson
loops constructed by Zarembo \cite{Zarembo:2002an}.

It was shown in \cite{Drukker:2006ga} that up to order $(g_{YM}^2 N)^2$
all interacting graphs in the Feynman gauge cancel and the only
contribution comes from ladder diagrams where the propagator is a
constant proportional to $\cos^2\theta_0$. This naturally led to
the conjecture that the expectation value of this Wilson loop is given by the same matrix
model as the half-BPS one \cite{Erickson:2000af,Drukker:2000rr}
with the replacement of the coupling $\lambda$ by
$\lambda'=\lambda\cos^2\theta_0$. 
This gives the prediction
\begin{equation}
\vev{W_{\theta_0}}
=\frac{1}{N}L_{N-1}^1\left(-\frac{\lambda'}{4N}\right)
\exp\left[\frac{\lambda'}{8N}\right]\,,
\end{equation}
where $L_{N-1}^1$ is a Laguerre polynomial. In \cite{Drukker:2006ga}
only the planar limit of this expression
\begin{equation}
\vev{W_{\theta_0,\,\text{planar}}}
=\frac{2}{\sqrt{\lambda'}}I_1\left(\sqrt{\lambda'}\right)\,,
\label{bessel}
\end{equation}
was considered ($I_1$ is a modified Bessel function).
String theory provided exact agreement with the strong
coupling expansion of this expression, as we shall review shortly.
Furthermore, the same rescaling was observed in the 
computation of correlation functions between this $1/4$ BPS loop and 
chiral primary operators \cite{Semenoff:2006am}.

In the present calculation we want to capture a different limit,
beyond the planar one. We consider a multiply wrapped Wilson loop,
or a loop in the $k$-th symmetric representation\footnote{It has been
proven in \cite{Okuyama:2006jc} and \cite{Hartnoll:2006is} 
that in the matrix model the multiply wound loop $W^{(k)}$ and the 
totally symmetric operator $W_{S_k}$ coincide in the strong coupling
regime.}, keeping the quantity $\kappa'\equiv k\sqrt{\lambda'}/4N$
fixed while taking both $N$ and $\lambda$ to infinity. This is the
limit that was discussed in
\cite{Drukker:2005kx,Yamaguchi:2006tq,Okuyama:2006jc,Hartnoll:2006is}
for the half-BPS loop, and in this limit the matrix model reduces
to
\begin{equation}
\vev{W_{\kappa'}} =\exp\left[2N\left (\kappa'\sqrt{1+\kappa'^2}
+\text{arcsinh}\,\kappa'\right)\right]\,.\label{MM1/4}
\end{equation}
There is also a subleading contribution, that we did not include in the 
formula above, obtained by replacing $\kappa'\rightarrow -\kappa'$. 
The appearance of this term can be explained by the fact that perturbation 
theory should be invariant under 
$\lambda'\rightarrow e^{2 i \pi}\lambda'$.  At strong coupling the 
expectation value of the Wilson loop depends on $\sqrt{\lambda'}$, 
so that an extra term with $\kappa'\rightarrow -\kappa'$ is needed. In 
the planar approximation this subleading term reduces to 
$e^{-\sqrt{\lambda'}}$, which 
appears in the large $\lambda'$ expansion of the Bessel function in 
(\ref{bessel}).

Later in this section we will be able to construct a D3-brane which
is dual to the multiply wrapped 1/4 BPS Wilson loop and we will
recover (\ref{MM1/4}) from supergravity. This computation will also produce the subleading contribution 
discussed above, which will correspond to an unstable D3-brane solution.

\subsection{String solution}

To write the relevant string solutions in the dual supergravity picture we
use the following metric on $AdS_5\times S^2$ (as in the previous example we drop
the directions on $S^5$ which do not play a role here)
\begin{equation}
\frac{ds^2}{L^2}=
-d\chi^2+\cos^2\chi(d\rho^2+\sinh^2\rho\,d\psi^2)
+\sin^2\chi(d\sigma^2+\sinh^2\sigma\,d\varphi^2)
+d\theta^2+\sin^2\theta d\phi^2\,.
\label{metricH2xH2}
\end{equation}
This metric has Lorentzian signature, which is somewhat more natural for
the supersymmetry analysis, but later we will also use the
Euclidean version obtained by Wick rotating $\chi\to iu$ and
$\sigma\to i\vartheta$
\begin{equation}
\frac{ds^2}{L^2}=
du^2+\cosh^2 u(d\rho^2+\sinh^2\rho\,d\psi^2)
+\sinh^2 u(d\vartheta^2+\sin^2\vartheta\,d\varphi^2)
+d\theta^2+\sin^2\theta d\phi^2\,.
\label{metric-H2xS2}
\end{equation}
Note that in the Lorentzian case the $\chi$ coordinate foliates $AdS_5$
by $\bH_2\times\bH_2$ surfaces ($\bH_2$ is the two-dimensional 
hyperbolic space, or Euclidean $AdS_2$), while in the Euclidean case $u$ 
foliates it into $\bH_2\times S^2$ surfaces.

The string describing the Wilson loop (\ref{WL-circle}) will be at 
$\chi=0$ (or $u=0$) and should end at $\rho\to\infty$ along 
a circle  parameterized by $\psi$. As we go along this circle we should 
also move along a circle on $S^2$, the parallel at angle $\theta_0$ 
spanned by the angle $\phi$. We take the ansatz where along the entire 
world-sheet we equate $\psi$ and $\phi$. As mentioned, the asymptotic 
value of $\theta$ should be $\theta_0$.
In \cite{Drukker:2006ga} two solutions with these boundary
conditions were found
\begin{equation}
\phi=\psi\,,\qquad
\sinh\rho(\sigma)=\frac{1}{\sinh\sigma}\,,\qquad
\sin\theta=\frac{1}{\cosh(\sigma_0\pm\sigma)}\,.\label{stringsol0}
\end{equation}
Here $\sigma$ is a world-sheet coordinate and  $\sigma_0$ is
related to the boundary value of $\theta$ by
\begin{equation}
\sin\theta_0=\frac{1}{\cosh\sigma_0}\,.
\end{equation}
One can eliminate $\sigma$ from the previous equations to find the
relation
\begin{equation}
\cosh\rho \cos\theta \sin\theta_0 - \sinh\rho \sin\theta
\cos\theta_0 = \pm \sin\theta_0.\label{stringsol}
\end{equation}
The two sign choices correspond to surfaces extending over the north
and south pole of $S^2$ respectively. The classical action for the
two cases is equal to
\begin{equation}
S=\mp\cos\theta_0\sqrt\lambda=\mp\sqrt{\lambda'}\,.
\end{equation}
The dominant contribution has negative action and corresponds to
the surface extended over less than half a sphere. That solution
is stable, while the one extending over the other pole has
positive action and three unstable modes.

These two solutions were interpreted in \cite{Drukker:2006ga} as
corresponding to the two saddle points in the asymptotic expansion of
the Bessel function (\ref{bessel})
\begin{equation}
\vev{W_{\theta_0,\,\text{planar}}}
\mathop{\longrightarrow}_{\lambda'\to\infty}
\frac{\sqrt{2}}{\sqrt{\pi}\lambda'^{3/4}}
\left[
e^{\sqrt{\lambda'}}\left(1+\mathcal{O}(1/\sqrt{\lambda'})\right)
-ie^{-\sqrt{\lambda'}}\left(1+\mathcal{O}(1/\sqrt{\lambda'})\right)\right]\,.
\end{equation}
Furthermore, it was shown there that considering the limit of large
$\lambda$, while keeping small $\lambda'$ and integrating over the
three modes that are massless for $\lambda'=0$, yields an identical result
to the full planar expression from the matrix model (\ref{bessel}), including 
all $\alpha'$ corrections.

The counting of the supersymmetries for the
solutions (\ref{stringsol0}) goes
very similarly to the counting presented in Section 2 for the loop with
insertions. 
The dependence of the Killing spinors on the relevant components of the metric (\ref{metricH2xH2}) is\footnote{
This is presented in greater detail in the next subsection (\ref{killing}) 
together with the vielbeins (\ref{vielbeins}).}
\begin{equation}
\epsilon=e^{-\frac{i}{2}\rho\,\Gamma_\star\Gamma_1}
e^{\frac{1}{2}\psi\,\Gamma_{12}}
e^{-\frac{i}{2}\theta\,\Gamma_\star\Gamma_5}
e^{\frac{1}{2}\phi\,\Gamma_{56}}\epsilon_0\,,
\label{Killing-string-sec3}
\end{equation}
while the constraints analogous to (\ref{SUSY-cond-1}) and (\ref{SUSY-cond-2}) 
are now
\begin{equation}
(\Gamma_{12}+\Gamma_{56})\epsilon_0=0\,, \label{SUSY-cond-12}
\end{equation}
and 
\begin{equation}
K\epsilon_0 = -\Big(\cos\theta_0\,\Gamma_{12}
+\sin\theta_0\Gamma_{16}\Big)\epsilon_0\,.\label{proj2}
\end{equation}
The two conditions (\ref{SUSY-cond-12}) and (\ref{proj2}) are
compatible and therefore the two string solutions
(\ref{stringsol0}) preserve one quarter of the supersymmetries, as
does the operator $W_{\theta_0}$ in the dual gauge theory.

\subsection{D3-brane solution}

We now move on to the construction of the 1/4 BPS D3-brane which 
describes the 
circular Wilson loop in the $k$-th symmetric representation $W_{\kappa'}$.
The supersymmetry analysis will be presented in 
Lorentzian signature (\ref{metricH2xH2}) to avoid defining 
the Killing spinors in Euclidean space. The resulting brane 
has extra factors of $i$ in the projector equations and 
an over-critical electric field. Moreover 
it does not seem to correspond to a Wilson loop operator in the gauge theory, 
but to a higher-dimensional observable. Still we 
find this way of performing the calculation useful. 
After presenting the solution we will switch to Euclidean signature 
(\ref{metric-H2xS2}), where the solution will not suffer from those 
problems and will be perfectly well defined.

We parameterize the brane world-volume 
by $\{\rho,\, \psi,\, \sigma,\, \varphi\}$.
The 1/2 BPS brane has constant $\chi=\arcsin\kappa$ and $\theta=0$. 
A natural ansatz for the 1/4 BPS brane is then to take 
$\chi=\chi(\rho)$, $\theta=\theta(\rho)$ and identify $\psi$ 
with $\phi$. This is consistent with the symmetries of the loop.
The asymptotic value of $\theta$ at
$\rho=\infty$ should be $\theta_0$. To carry the $k$ units of flux 
represented by the Wilson loop operator, we switch on an electric field
$F_{\rho\psi}(\rho)$.

Note that the dependence of $\theta$  and $\chi$ 
on $\rho$ explicitly breaks the $AdS_2$
isometry. This fact makes it difficult to guess a simple ansatz for
the solution to the equations of motion of the brane.
 As in Section 2 we will then proceed by looking at the first-order
supersymmetry equations which follow from requiring $\kappa$-symmetry 
on the brane world-volume. 

We begin by constructing the Killing spinors associated to the
$AdS_5\times S^2$ metric (\ref{metricH2xH2}) with Lorentzian signature. The vielbeins
relevant for the D3-brane solution are
\begin{equation}
\begin{gathered}
e^0=L\,d\chi\,,\qquad 
e^1=L\cos\chi\,d\rho\,,\qquad
e^2=L\cos\chi\sinh\rho\,d\psi\,,
\\
e^3=L\sin\chi\,d\sigma\,,\qquad
e^4=L\sin\chi\sinh\sigma\,d\varphi\,,\qquad
\\
e^5=L\,d\theta\,,\qquad
e^6=L\sin\theta\,d\phi\,.\qquad
\end{gathered}
\label{vielbeins}
\end{equation}
Using the same notation of Section 2, the Killing spinors may then be written as
(adding the dependence on $\vartheta$ and $\varphi$ to 
(\ref{Killing-string-sec3}))
\begin{equation}
\epsilon=e^{-\frac{i}{2}\chi\Gamma_*\Gamma_0}
e^{-\frac{i}{2}\rho\Gamma_*\Gamma_1}
e^{\frac{1}{2}\psi\Gamma_{12}} e^{-\frac{1}{2}\sigma\Gamma_{03}}
e^{\frac{1}{2}\varphi\Gamma_{34}}
e^{-\frac{i}{2}\theta\,\Gamma_\star\Gamma_5}
e^{\frac{1}{2}\phi\,\Gamma_{56}}\epsilon_0\,.
\label{killing}
\end{equation}

The DBI Lagrangian reads (with the sign in the square root appropriate for a 
brane with Euclidean world-volume and with $F_{\rho\psi}$ containing a 
factor of $2\pi\alpha'/L^2$)
\begin{equation}
\mathcal{L}_{DBI}
=L^4 
\sin^2\chi
\sinh\sigma \sqrt{(
-\chi'^2+\theta'^2+\cos^2\chi)(\cos^2\chi\sinh^2\rho+\sin^2\theta)
+F^2_{\rho\psi}}\,,
\end{equation}
and the projector associated with the D3-brane is
\begin{equation}
\Gamma=\cL_{DBI}^{-1}\left(i\Gamma_{(4)}-L^2F_{\rho\psi}\Gamma_{(2)}K\right)I\,,
\end{equation}
where, again,  $K$ acts by complex conjugation, $I$ by multiplication by $-i$
and
\begin{equation}
\begin{aligned}
\Gamma_{(4)}&= (\gamma_\rho+\chi'\,\gamma_\chi+\theta'\gamma_\theta)
(\gamma_\psi+\gamma_\phi)\,\Gamma_{(2)}\\
&=L^2 (\cos\chi\Gamma_1+\chi'\Gamma_0+\theta'\Gamma_5)
(\cos\chi\sinh\rho\Gamma_2+\sin\theta\Gamma_6)\,\Gamma_{(2)}
\\
\Gamma_{(2)}&=
\gamma_\sigma\gamma_{\varphi}=L^2 \sin^2\chi\sinh\sigma\,\Gamma_{34}\,.
\end{aligned}
\end{equation}

Note that the projector $\Gamma$ does not depend on $\psi$. As for the string case we can
eliminate the dependence on $\psi$ in the projection equation by
imposing
\begin{equation}
(\Gamma_{12}+\Gamma_{56})\epsilon_0=0\label{Gamma12Gamma56}\,,
\end{equation}
and then we impose also the condition
\begin{equation}
K\epsilon_0=-(\cos\theta_0\Gamma_{12}
+\sin\theta_0\Gamma_{16})\epsilon_0\, , 
\label{gamma1256}
\end{equation}
which both follow from the analysis of the supersymmetries of the
string (\ref{proj2}). The brane solution will then preserve 
the same quarter of supersymmetries as the string and as the gauge theory 
observable.

Because of the isometry of the system the factor of
\begin{equation}
M\equiv e^{-\frac{1}{2}\sigma\Gamma_{03}}e^{\frac{1}{2}
\varphi\Gamma_{34}}
\end{equation}
commutes with those two constraints, so $\epsilon_0$ and $M\epsilon_0$ 
satisfy the same conditions.

Using some gamma-matrix algebra and applying the
constraints above, we move the matrix
$e^{-\frac{i}{2}\chi\,\Gamma_\star\Gamma_0}e^{
-\frac{i}{2}\rho\,\Gamma_\star\Gamma_1}
e^{-\frac{i}{2}\theta\,\Gamma_{\star}\Gamma_5}$ to the left of
$\Gamma$ in the projection equation. In this way one gets a set of
8 first order differential equations for $\theta$, $\chi$ and $F_{\rho\psi}$
(indicating which gamma-matrix combination leads to them)
\begin{equation}
\begin{array}{lcl}
\Gamma_{0234}:
&&0=iF_{\rho\psi}\sin\chi\sin\theta\sin\theta_0
+\chi'\sinh\rho(\cos^2\chi-\sin^2\theta)+
\cosh\rho\sin\chi\cos\chi\sin^2\theta\,
\cr 
\Gamma_{\star} \Gamma_5:
&&0=iF_{\rho\psi}\sin\chi\sin\theta\cos\theta_0
-\chi'\cosh\rho\sin\theta\cos\theta+
\sinh\rho\sin\chi\cos\chi\sin\theta\cos\theta\,
 \cr
\Gamma_{1234}:
&&0=iF_{\rho\psi}\sinh\rho\cos\chi\sin\theta\sin\theta_0
+iF_{\rho\psi}\cosh\rho\cos\chi\cos\theta\cos\theta_0-
\cr &&\hskip .7cm -\chi' \sinh^2\rho\sin\chi\cos\chi-\theta'
\sin\theta\cos\theta+\sinh\rho\cosh\rho\cos^2\chi\,
\cr
\Gamma_{2345}: 
&&0=iF_{\rho\psi}\sinh\rho\cos\chi\sin\theta\cos\theta_0
-iF_{\rho\psi}\cosh\rho\cos\chi\cos\theta\sin\theta_0-
\cr &&\hskip .7cm -\theta'
\sinh\rho\cosh\rho\cos^2\chi-\cos^2\chi\sin\theta\cos\theta\,
\cr
\Gamma_{01}:
&&0=iF_{\rho\psi}\sinh\rho\cos\chi\cos\theta\cos\theta_0
+iF_{\rho\psi}\cosh\rho\cos\chi\sin\theta\sin\theta_0-
\cr &&\hskip .7cm -\chi' \sinh\rho\cosh\rho\sin\chi\cos\chi
+\cos^2\chi(\sinh^2\rho+\sin^2\theta)\,
\cr
\Gamma_{05}:
&&0=iF_{\rho\psi}\sinh\rho\cos\chi\cos\theta\sin\theta_0
-iF_{\rho\psi}\cosh\rho\cos\chi\sin\theta\cos\theta_0\\
&&\hskip.7cm
+\theta'(\sinh^2\rho\cos^2\chi+\sin^2\theta)\,
\cr
\Gamma_{15}: 
&&0=iF_{\rho\psi}\sin\chi\cos\theta\sin\theta_0
-\chi'\sinh\rho\sin\theta\cos\theta+\cr
&&\hskip .7cm +\theta'\sinh\rho\sin\chi\cos\chi+
\cosh\rho\sin\chi\cos\chi\sin\theta\cos\theta\,
\cr
1:
&&1=-i L^4 \cL_{DBI}^{-1}\sin^2\chi\sinh\sigma
\left(iF_{\rho\psi}\sin\chi\cos\theta\cos\theta_0+
\chi'\cosh\rho\sin^2\theta
\right.\\&&\hskip.7cm\left.
-\sinh\rho\sin\chi\cos\chi\sin^2\theta\right)\,
\label{projequations}
\end{array}
\end{equation}
The half-BPS solution
\begin{equation}
\chi'=0\, , \qquad
F_{\rho\psi}=i\cos\chi \sinh\rho
\end{equation}
can be recovered by setting $\theta=\theta_0=0$ in (\ref{projequations}). 

The equations (\ref{projequations}) are all consistent with each
other. One can solve any three of them, the remaining ones being
automatically satisfied.
The first three, for example, lead to the equations
\begin{eqnarray}
&&\theta'=A\,\cos^2\chi\cos\theta\, ,
\qquad
\chi'=A\, \sin\chi\cos\chi\sin\theta\, ,  \\
&&F_{\rho\psi}=-i\frac{\cos\chi\cos\theta}{\cos\theta_0} (A\,
\cosh\rho\sin\theta -\sinh\rho)\, ,
\end{eqnarray}
where
\begin{equation}
A=\frac{\sinh\rho\cos\theta\sin\theta_0
-\cosh\rho\sin\theta\cos\theta_0}
{(\cos^2\chi-\sin^2\theta)\sinh\rho\cos\theta_0
+\cosh\rho\sin\theta\cos\theta\sin\theta_0}\, .
\end{equation}
Taking the ratio of $\theta'$ and $\chi'$ yields
\begin{equation}
\sin\chi \cos\theta = C\, ,
\label{chi-sol}
\end{equation}
where $C$ is an integration constant. Inserting this solution into the expression for 
$\theta'$ and solving the resulting differential equation gives
\begin{equation}
\cos\chi \left ( \cosh\rho \cos\theta \sin\theta_0 - \sinh\rho
\sin\theta \cos\theta_0 \right ) = D\, . \label{theta_sol}
\end{equation}

On the $AdS_5$ side of the ansatz at $\rho=0$ the circle parameterized 
by $\psi$ shrinks to a point. For the solution not to be singular at that 
point, the same has to happen also on the $S^5$ side, since $\phi=\psi$. 
The solution will be regular at $\rho=0$ only if at that point 
$\sin\theta=0$, which then gives $D$ in terms of $\theta_0$ and $C$ as
\begin{equation}
D=\pm \sin\theta_0 \sqrt{1-C^2}\, ,
\end{equation}
where the $+,\, -$ signs correspond respectively to taking either $\theta=0$ or
$\theta=\pi$ at $\rho=0$, or, in other words, to
wrapping the brane around the northern or the southern hemisphere of $S^2$.
Notice that in the string limit ($\chi \rightarrow 0$, or $C
\rightarrow 0$) the expression (\ref{theta_sol}) reduces to
 the string solution (\ref{stringsol}).

These solutions in Lorentzian space are unphysical. The world-volume 
of the brane is Euclidean, but the electric field is over-critical, leading to 
an imaginary action. Furthermore, the branes do not end along curves on 
the boundary, but along higher-dimensional surfaces, and do not provide 
a holographic description of Wilson loops.

Therefore we analytically continue those solutions to Euclidean signature, 
where the resulting branes will provide a good holographic dual of Wilson 
loop operators. We take the Wick rotation
\begin{equation}
\chi= i\,u\, , \qquad
 \sigma= i\, \vartheta\,.
\end{equation}
In these coordinates, the Euclidean $AdS_5$ is written as an 
$\bH_2\times S^2$ fibration
as in eq. (\ref{metric-H2xS2}).
The solution (\ref{chi-sol}) and (\ref{theta_sol}) in Lorentzian 
signature now becomes
\begin{equation}
\sinh u \cos\theta = c\, ,
\label{u-sol}
\end{equation}
and
\begin{equation}
\cosh u \left ( \cosh\rho \cos\theta \sin\theta_0 - \sinh\rho
\sin\theta \cos\theta_0 \right ) = d\, .\label{rho_sol}
\end{equation}
Similarly to the Lorentzian case the solution is smooth at $\rho=0$ only 
for
\begin{equation}
d=\pm \sin\theta_0 \sqrt{1+c^2}\,.
\end{equation}
The implicit equation (\ref{rho_sol}) is solved for $\rho$ as a function 
$\theta$ by
\begin{eqnarray}
\sinh\rho&=&\mbox{sign}(\theta_0-\theta)
\frac{\sin\theta\sin\theta_0\left(\sqrt{1+c^2}\cos\theta_0+
\cos\theta\sqrt{1+\frac{c^2\cos^2\theta_0}{\cos^2\theta}}\right)}{\cosh
u (\cos^2\theta-\cos^2\theta_0)}\cr
\cosh\rho&=&\mbox{sign}(\theta_0-\theta)
\frac{\sqrt{1+c^2}\cos\theta\sin^2\theta_0+
\sin^2\theta\cos\theta_0\sqrt{1+\frac{c^2\cos^2\theta_0}{\cos^2\theta}}}{\cosh
u(\cos^2\theta-\cos^2\theta_0)}\,.\cr && \label{rhooftheta}
\end{eqnarray}
The sign function allows us to write in a single expression the two solutions 
corresponding to a brane wrapping over the north or south poles of 
the $S^2$. We will assume, without loss of generality, that 
$\theta_0\leq\pi/2$.

Given that the solution may be written explicitly as a function of 
$\theta$, it makes sense to use it, instead of $\rho$, as one of the 
world-volume coordinate. Thus the world-volume is parameterized 
by $\{\theta,\, \psi,\, \vartheta,\, \varphi\}$ and 
$\rho=\rho(\theta)$ and $u=u(\theta)$ are given by the 
solutions above. This parametrization will be singular in the 
$1/2$ BPS limit, where $\theta=0$, but that solution is very simple, 
with arbitrary $\rho$ and constant $u=\text{arcsinh}\,c$.

The DBI action in this signature reads
\begin{equation}
S_{DBI} = 4N \int d\theta\, \sinh^2 u \, 
\sqrt{(\cosh^2 u\,\rho'^2 + u'^2 +1)(\cosh^2 u \sinh^2 \rho + 
\sin^2\theta)+F^2_{\theta\psi}}\, ,
\label{DBIeucl}
\end{equation}
while the Wess-Zumino term can be written as
\begin{equation}
S_{WZ}= 4N \int d\theta\, \rho' \sinh\rho \left( \frac{u}{2}
-\frac{1}{2}\sinh u \cosh u - \sinh^3 u \cosh u \right)\,. \label{WZeucl}
\end{equation}
To obtain these expressions we have integrated over $\psi$ and
$S^2$. Now the $'$ stands for the derivative with respect to
$\theta$. We have checked that the solutions found above satisfy the equations of motion coming from (\ref{DBIeucl}) and (\ref{WZeucl}).

\begin{figure}
\begin{center}
\includegraphics[width=90mm]{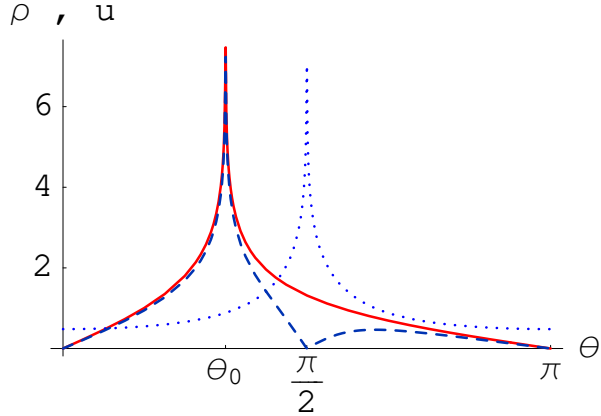}
\caption{A depiction of string and D3-brane solutions. 
The solid line gives $\rho$ as a function of $\theta$ for the string 
with boundary value of $\theta=\pi/3$. The D3-brane solution is 
represented by the dashed and dotted lines which are respectively 
$\rho$ and $u$ as functions of $\theta$ (for $\kappa=1$). 
In both cases there are two solutions, a stable one with 
$0\leq\theta\leq\theta_0$ (where $\rho$ for the string 
and D3-brane are nearly indistinguishable) and an unstable one, with 
$\theta\leq\theta_0\leq\pi$. The unstable D3-brane solution 
reaches the boundary of $AdS$ not only at $\theta_0$, but also 
at $\theta=\pi/2$, where $u$ diverges, but then it turns back and 
closes smoothly on itself.}
\label{plotrho}
\end{center}
\end{figure}

In Figure~\ref{plotrho} we have plotted 
$\rho$ and $u$ as  functions of $\theta$ for a D3-brane solution 
and for comparison also $\rho$ for the analog string solution (in which 
case $u=0$). 
There are two solution, both reaching infinite $\rho$ at 
$\theta_0$ (for the example pictured we took the values 
$\theta_0=\pi/3$ and $\kappa=1$). The stable 
solution then goes to $\rho=0$ at $\theta=0$, while the unstable 
solution goes to $\rho=0$ at $\theta=\pi$.

Note that in the case of the unstable D3-brane solution the coordinate $u$ 
diverges at the equator $\theta=\pi/2$, as can also be seen from 
(\ref{u-sol}). This means that the D3-brane reaches the boundary 
of $AdS$ at that point, and gets reflected back into the interior (after changing the sign of $u$). One 
could choose to truncate the surface there and consider either half of 
the solution. But in the dual gauge theory that will not correspond to a 
Wilson loop vacuum expectation value. Rather, the D3-brane extending 
from $\theta_0$ to $\theta=\pi/2$ will be the correlator between 
the Wilson loop and a two-dimensional surface operator 
located where the brane reaches the boundary (the surface spanned by 
$\{\vartheta,\,\varphi\}$, the radius of the $\psi$ circle shrinks 
to a point there). The other part of the solution, from 
$\theta=\pi/2$ to $\theta=\pi$ is the vacuum expectation value 
of the surface operator itself, with no Wilson loop insertion.

As usual the BPS equations simplify the square root in the DBI
action, which in this case reduces to
\begin{equation}
S_{DBI}=
 4N\int d\theta \left |F_{\theta\psi}  \frac{\cos\theta_0}{\cos\theta}\right| \sinh^3 u \,. \label{ldbi}
\end{equation}
This fact can be used to check the conservation of $\Pi$, the
momentum conjugate to the gauge field $A_\psi$
\begin{equation}
\Pi = -i\,\frac{2\pi \alpha'}{L^2} \, T_{D3}
\int d\vartheta\, d\varphi \frac{\delta
{\cal L}_{DBI}}{\delta F_{\theta\psi}} = \pm
\frac{4N}{\sqrt{\lambda}}\left| \frac{c}{\cos\theta_0} \right|
\equiv \pm k,
\end{equation}
where the two signs correspond to the two solutions. This implies
that
\begin{equation}
|c| =  \kappa |\cos\theta_0|.
\label{c}
\end{equation}

To compute the full on-shell action we have to supplement the DBI
and WZ bulk contributions with total derivatives and 
boundary terms associated to the
electric field $F_{\theta\psi}$ and the scalar $\rho(\theta)$.
These are given respectively by\footnote{For a general discussion
on the role of boundary terms see
\cite{Drukker:1999zq} and \cite{Drukker:2005kx}.}
\begin{equation}
S_{L.T.}=-  i\, \frac{L^2}{2\pi\alpha'} \int d \theta\, d\psi\,  \Pi\,  F_{\theta \psi}\,,
\end{equation}
and
\begin{equation}
-P_{\rho}\Big{|}_{\theta \rightarrow \theta_0} = -T_{D3}\int d\psi\,
d\vartheta\, d\varphi\, \frac{\delta ({\cal L}_{DBI}+{\cal
L}_{WZ})}{\delta \rho'}\bigg |_{\theta \rightarrow
\theta_0}\label{prho}.
\end{equation}
where $P_{\rho}$ is the momentum conjugate to $\rho$.

The boundary term for $\rho$ can be motivated as follows.  Let us
consider the $AdS_5$ metric in the Poincar\'e patch
\begin{equation} 
ds^2=\frac{1}{z^2}
(dz^2+dr_1^2+r_1^2d\varphi_1^2+dr_2^2+r_2^2d\varphi_2^2)\,.
\end{equation}
The transformation relating $z$ to our coordinates is
\begin{equation}
z=\frac{1}{ \cosh u\cosh\rho-\cos\vartheta\sinh u}\,,
\label{transfor}
\end{equation}
so that the $\rho\rightarrow \infty$ region corresponds to $z=0$.
In the Poincar\'e patch the boundary term associated to $z$ has the
form of a Legendre transform evaluated at the  boundary of $AdS_5$
\cite{Drukker:1999zq}\cite{Drukker:2005kx}
\begin{equation}
-\int \,z\, p_z\Big{|}_{z\sim 0}\,.
\end{equation}
Using (\ref{transfor}) it is immediate to verify that in
proximity of the boundary $z\, p_z\sim p_{\rho}$. This
justifies the form of the boundary term for $\rho$.

 Now we can evaluate the on-shell action. The bulk and boundary contributions diverge as
we approach the boundary of $AdS_5$, {\it i.e.} in the limit
$\theta\rightarrow\theta_0$. We can regularize these divergences
by introducing a cut-off at $\theta_0-\epsilon$. 
This leads to the following expression for the regularized DBI action
\begin{eqnarray}
S_{DBI}&=&T_{D3} \int_0^{\theta_0-\epsilon}d\theta\int d\psi\,
d\vartheta\, d\varphi\,{\cal L}_{DBI}=4N \kappa^3\sin{\theta_0}\frac{\sqrt{1+c^2}}{\epsilon}\nonumber\\
&&+2N\kappa^3\, \frac{\sin{\theta_0}\tan{\theta_0}}{\sqrt{1+c^2}}
-N\kappa\sec^2{\theta_0}(2+8 c^2-4
c^2\cos^2{\theta_0})\sqrt{1+c^2
\cos^2{\theta_0}}\nonumber\\
&&+2N\kappa
\sec^3{\theta_0}\sqrt{1+c^2}\,\,\left(1+3c^2-6c^2
\cos^2{\theta_0}+2c^2
\cos^4{\theta_0}\right)\nonumber\\
&&-2N\sec^4{\theta_0}\left(1-4c^2\sin^2{\theta_0}-8c^4\sin^2{\theta_0}\right)\log\left(
\frac{{c+\sqrt{1+c^2}}}{c\,\cos{\theta_0}+\sqrt{1+c^2\cos^2{\theta_0}}}\right)\nonumber\\
&&+8N\kappa^3\sin{\theta_0}\tan{\theta_0}\sqrt{1+c^2}\Big{(}2\log{\epsilon}-2\log(\cos{\theta_0}\sin{\theta_0})-\log\left({1+c^2}\right)\nonumber \\
&&+\log\Big{(}\cos^2{\theta_0}\left(1+2 c^2\right)+2
\cos{\theta_0}\sqrt{1+c^2}\sqrt{1+c^2\cos^2{\theta_0}}+1\Big{)}\Big)\,
.
\end{eqnarray}
The Legendre transform of the gauge field is written as the integral over the 
total derivative
\begin{eqnarray}
S_{L.T.}&=&-i\,  \frac{L^2}{2\pi\alpha'}\int_0^{\theta_0-\epsilon}d\theta\int d\psi\,
 \Pi\,  F_{\theta
\psi}\nonumber\\&=&4N\kappa \sin{\theta_0}
\frac{\sqrt{1+c^2}}{\epsilon}-2N c
\frac{3+\kappa^2+2c^2}{\sqrt{1+c^2}}\,,
\end{eqnarray}
and the boundary term for $\rho$ is
\begin{eqnarray}
-P_\rho\Big{|}_{\theta_0-\epsilon}&=&-2N\sin{\theta_0}(\kappa\,\sqrt{1+\kappa^2}+\text{arcsinh}\,\kappa)\,
\frac{\sqrt{1+c^2}}{\epsilon\,\,\sqrt{1+\kappa^2}}\nonumber\\
&&-4N\sec\theta_0(\kappa\,\sqrt{1+\kappa^2}+\text{arcsinh}\,\kappa)\,
\frac{2\kappa^2-(1+4\kappa^2-\kappa^4)\cos^2{\theta_0}-2c^4}{ 4(1+\kappa^2)^{3/2}\sqrt{1+c^2}}\, .\nonumber \\ 
\end{eqnarray}
Finally the regularized Wess-Zumino term turns out to be
\begin{equation}
{S}_{WZ}=-2N
\left(c\sqrt{1+c^2}+\text{arcsinh}\,c\right)-{S}_{DBI}-S_{L.T.}+P_\rho\,.
\end{equation}

Those expressions are much more complicated than the $1/2$ BPS case, 
where those three terms are
\begin{eqnarray}
S_{DBI}&=&4N \int d\rho\sinh\rho\cosh u\sinh^3u\,,
\nonumber\\
S_{WZ}&=& 4N \int d\rho \sinh\rho \left( \frac{u}{2}
-\frac{1}{2}\sinh u \cosh u - \sinh^3 u \cosh u \right)\,,
\\
S_{L.T.}&=&4N \int d\rho \sinh\rho \cosh u\, .\nonumber
\end{eqnarray}
The boundary term for $\rho$, just removes the divergence from the 
upper limit of $\rho$ integration, giving $-1$ from the lower limit. 
Summing up all the contributions and using $\sinh u=\kappa$ 
gives the full on-shell action \cite{Drukker:2005kx}
\begin{eqnarray}
S=-2N\left(\kappa\sqrt{1+\kappa^2}+\text{arcsinh}\,\kappa\right)\,.
\end{eqnarray}

While the regularized expressions for the $1/4$ BPS loop are much more 
complicated, the sum of the bulk and boundary terms is exactly the 
same with the replacement of $\kappa$ by $c$
\begin{equation}
S_\text{total}= - 2N \left(c \sqrt{1+c^2}+\text{arcsinh}\, c\right).
\end{equation}
Recall that $c$ is related to the number of units of flux carried by the 
brane, or the rank of the representation of the Wilson loop by 
(\ref{c})
\begin{equation}
c=\kappa'=\frac{k\cos\theta_0\sqrt\lambda}{4N}\,.
\end{equation}
Therefore this stable solution will contribute to the expectation value of the 
$1/4$ BPS Wilson loop at strong coupling
\begin{equation}
\vev{W_{\kappa'}} =\exp\left[2N\left
( \kappa'\sqrt{1+ \kappa'^2} +\text{arcsinh}\, \kappa'\right)\right]\,.
\end{equation}
This is the same result as can be derived from the matrix model observable 
(either the multiply wrapped loop \cite{Drukker:2005kx} or the 
symmetric one \cite{Okuyama:2006jc,Hartnoll:2006is}) in this limit. 
This serves as a confirmation that the matrix model correctly captures 
the $1/4$ BPS loop including all $1/N$ corrections at large $\lambda$.

An analogous computation can be done for the unstable branch, where the 
range of integration for
the coordinate $\theta$ is $[\theta_0+\epsilon,\,\pi]$ and $P_\rho$
is evaluated at $\theta_0+\epsilon$. The final result is exactly as above, 
except for the overall sign
\begin{equation}
S_\text{total}^\text{(unstable)}= 2N \left( \kappa' \sqrt{1+ \kappa'^2} + \text{arcsinh} \, \kappa'\right)\,.
\end{equation}
As in the case of the string solution reviewed before, this should correspond 
to an exponentially small correction to the expectation value of the Wilson 
loop when doing the asymptotic expansion at large $N$ and large $\lambda$ (see the discussion after eq. (\ref{MM1/4})).

In addition to the $1/2$ BPS limit, with $\theta_0=0$, there is another 
interesting limiting case, of $\theta_0=\pi/2$ studied by Zarembo 
\cite{Zarembo:2002an}.  In that case the two D3-brane 
solutions, whose actions always have the opposite signs, are degenerate. 
Both have vanishing action, and in fact there are more than two solutions, 
rather a whole family parameterized by an $S^3$. But, unfortunately, 
looking at the solutions 
at this limit we find that they do not provide a good description for the 
Wilson loop. If we consider finite $\kappa$, then from (\ref{c}), 
the constant $c$ vanishes and by (\ref{u-sol}), also $u=0$. 
Therefore the D3-brane 
shrinks to a two-dimensional surface and therefore the higher-derivative 
corrections to the DBI action cannot be ignored.

If instead we keep $c$ finite in that limit, 
then $\kappa$ will diverge, leading to a smooth D3-brane solution. But 
now as $\theta$ goes to $\theta_0$ both $\rho$ and $u$ diverge, meaning 
that the brane ends along a 3-dimensional surface on the boundary, 
rather than the Wilson loop.

\section{Discussion}

We have presented some solutions for D3-branes in $AdS_5\times S^5$, 
which are dual to certain $1/4$ BPS Wilson loop operators in $\cN=4$ 
supersymmetric Yang-Mills theory. The first example was a combined 
system of a loop with two local insertions made from complex scalar 
fields. Without the insertions the loop itself would have been $1/2$ BPS 
and the trace of the local insertions is also $1/2$ BPS, while the combined 
system preserves $1/4$ of the supersymmetries. The second system 
was a family of Wilson loops with couplings to three of the scalars 
in a way that also preserves eight supercharges.

It is by now a standard feature of the $AdS$/CFT correspondence that 
very long operators in the gauge theory map to ``giant'' D-brane objects 
rather than to fundamental strings or supergravity modes. 
In our case the D3-branes should 
describe the Wilson loops in a high-dimensional symmetric representation, 
where the rank of the representation $k$ is of order $N$. In the 
example of the loop with insertions we were able to calculate the 
energy and angular momentum and they agreed with each other, as 
would be expected, but there was no special feature arising from the 
fact that the loop is in a certain representation. 

In the second example we were able to compare the result of the 
$AdS$ calculation to a matrix model conjectured to describe those 
$1/4$ BPS loops 
\cite{Erickson:2000af,Drukker:2000rr,Drukker:2006ga}. 
The value of those loops at large $N$ and large 
$\lambda$ in a symmetric representation is known (and coincides 
with the single-trace multiply-wrapped loop). We found that the 
classical action for the D3-brane correctly reproduces the expected 
result, which includes an infinite series of $1/N$ corrections to the 
planar string expression. 
Furthermore, we have found two solutions with the same boundary 
conditions, in exact analogy with the strings describing the loop in the 
fundamental representation. The second solution, which contributes 
an exponentially small correction to the Wilson loop in the supergravity limit 
is the brane analog of a world-sheet instanton. Such contributions are 
expected, since the string expansion is asymptotic in $1/\sqrt\lambda$.

The geometry of this second D3-brane solution is very interesting. 
Starting from the boundary of $AdS$, where it originates along the 
Wilson loop, it moves into the bulk, turns back, goes again to the boundary, 
gets reflected back into the interior, and closes off smoothly on itself. If 
we chose not to continue the solution, it would end on a 
two-dimensional surface on the boundary. So this 
part of the solution would describe the correlator of a Wilson loop 
and a surface operator which are non-trivially linked.
 It would be very interesting to 
understand further the nature of this surface operator. The connection 
between Wilson loops and surface operators may not be so surprising 
given that they may both be described by branes in the bulk 
(see {\it e.g.} \cite{Constable:2002xt,Gomis:2006sb,Gukov:2006jk}).

As discussed at the end of the last section, one would like also to 
consider a special limit of these loops, when $\theta_0=\pi/2$. 
This limit is particularly interesting because of a comment made at the 
end of \cite{Drukker:2006ga}, where it was noticed that one may take 
$\lambda$ large while keeping $\lambda'=\lambda\cos\theta_0$ 
small, in a way similar to the BMN limit \cite{Berenstein:2002jq}. 
When considering the string solution in that limit, the mass of 
the string modes becomes much larger than the mass of the three 
broken zero modes (those parameterizing the $S^3$ mentioned above). 
Ignoring all the stringy modes and integrating only over those three 
leads to the full result of the planar matrix model, including all 
$\alpha'$ (or $1/\sqrt{\lambda'}$) corrections. It would be extremely 
interesting if we were able to repeat the calculation here and find the 
exact expression 
including all $1/N$ and $1/\sqrt{\lambda'}$ corrections. Recall that 
those corrections would not be the same for the loop in the symmetric 
representation and for the multiply-wound loop. So this calculation would 
be a very good check of the recent identification of the D3-brane with 
the loop in the symmetric representation \cite{Gomis:2006im}. 
Unfortunately, as explained before, in this limit the D3-brane degenerates 
and does not provide a good description of those Wilson loops.

The loops studied in this paper are not the most general $1/4$ BPS 
Wilson loops, all our examples had a circular geometry, which is not 
required. Many other loops were described in \cite{Zarembo:2002an}, 
and there are probably even more. The string solutions describing those 
loops were studied by Dymarsky et al. \cite{Dymarsky:2006ve}, and 
perhaps there is a general classification of the relevant branes along the 
lines of \cite{Mikhailov:2000ya}.

After studying the probe brane in the $AdS_5\times S^5$ background 
it is natural to consider the back-reaction of the brane on the geometry. 
This was pursued in the $1/2$ BPS case in \cite{Yamaguchi:2006te} 
and \cite{Lunin:2006xr}, where all the relevant metrics could 
be related to Young-tableaux, thus giving a correspondence 
between the representations of the Wilson loop and the associated metrics. 
It would be interesting to try to find the metrics in this case too, though 
this system has far less symmetry making it a much harder problem.

Finally, one can go further to a system which is only $1/8$ BPS, by 
looking at the correlators of those Wilson loops with chiral primary local 
operators. Amazingly, in the gauge theory those also seem to be captured 
fully by ladder diagrams and may be reduced to some matrix model 
\cite{Semenoff:2006am}. In the case of the $1/2$ BPS loop this was 
checked in $AdS$ using a string \cite{Berenstein:1998ij,Semenoff:2001xp}
and D-branes 
\cite{Giombi:2006de}. For the $1/4$ BPS loop this was done with 
a string in\cite{Semenoff:2006am} and would be interesting to 
repeat this calculation with D3-branes.

\subsection*{Acknowledgements}
We are happy to thank Nick Halmagyi, Igor Klebanov, Oleg Lunin, 
Niels Obers, Jan Plefka, Martin Ro\v{c}ek, Gordon Semenoff, Nemani Suryanarayana and Donovan Young for interesting conversations. SG is grateful for the hospitality of the 
Michigan Center for Theoretical Physics and of the Perimeter 
Institute at the final stage of this work. DT would like to thank 
Brown University, the University of British Columbia and the 
Enrico Fermi Institute at the University of Chicago for hospitality 
during the realization of this project. SG and DT acknowledge 
partial financial support through the NSF award PHY-0354776.

\end{document}